# Integrating Big Data and Survey Data for Efficient Estimation of the Median

Ryan Covey

Methodology and Data Science Division, Australian Bureau of Statistics

Level 3, 818 Bourke Street, Docklands, VIC 3000, Australia

ryan.covey@abs.gov.au

June 28, 2023

## Abstract

An ever-increasing deluge of big data is becoming available to national statistical offices globally, but it is well documented that statistics produced by big data alone often suffer from selection bias and are not usually representative of the population at large. In this paper, we construct a new design-based estimator of the median by integrating big data and survey data. Our estimator is asymptotically unbiased and has a smaller variance than a median estimator produced using survey data alone.

# 1 Introduction

The declining response rates and high respondent burden of surveys, alongside increasing budgetary pressures within national statistical offices and expanding availability of big data, are leading to high demand from official statisticians for methods that combine big data and survey data to produce affordable, accurate, representative and fit-for-purpose official statistics [1, 2]. Big data are often subject to selection bias that is catastrophic for estimation [3], and there is a growing literature on using surveys to correct this bias; see Rao [4] for a review.

We believe that this is the first paper to consider estimation of the median in this context. In Section 2, we define our design-based big- and survey-data median ('integrated median') alongside a median informed by the entire population ('population median') and the survey-data-only median ('survey median') of Kuk [5]. The integrated median is constructed from the integrated average of Kim and Tam [6] in the same way as the survey median is constructed from the Horvitz-Thompson average of Kuk [5]. Like Kim and Tam [6], we do not assume that the big data are missing-at-random [7], which is a strong and unverifiable assumption often made in this context [4]. In Section 3, all three medians are shown to converge to an ideal 'true' median ('superpopulation median') of an unknown superpopulation distribution that we assume generates the population. In Section 4, we compare each median on its asymptotic bias and variance as estimators of the superpopulation median, showing that while both the integrated and survey medians are asymptotically unbiased, the integrated median has a lower asymptotic variance if the big-data stratum is not empty. We finish with some concluding remarks and future research directions in Section 5 and relegate proofs of our results to the appendix in Section 8. In fact, our results hold for any quantile, not just the median (which is the 0.5-quantile), and these more general results are provided in the appendix; see Corollary 8.1 and Theorem 8.2.

## 2 Integrated Estimation

Consider a population $X_1, X_2, \ldots, X_n$ of independent and identically distributed (i.i.d.) observations drawn from an unknown superpopulation distribution, and suppose that we are interested in estimating the superpopulation median $\theta_0$, satisfying $\Pr(X_i \leq \theta_0) \geq 0.5$ and $\Pr(X_i \geq \theta_0) \geq 0.5$. Perhaps the most popular estimate is the population median

$$\hat{\theta}_n = \begin{cases} X_{((n+1)/2)} & n \text{ is odd} \\ (X_{(n/2)} + X_{(n/2+1)})/2 & n \text{ is even} \end{cases}, \qquad (1)$$

where $X_{(i)}$ is the $i^{\text{th}}$ smallest observation in the population, so that $X_{(1)} \leq X_{(2)} \leq \cdots \leq X_{(n)}$.

If obtaining the entire population is infeasible, we might want to conduct a probability survey and only sample a subset of observations $X_i$ for which $i$ lies in some randomly selected survey stratum $A$. Let $\alpha_i$ equal 1 or 0 according to whether or not $i$ lies in $A$, and let $d_i = \pi_i^{-1}$ be the design weight of unit $i$, where $\pi_i$ is the probability that unit $i$ is selected for inclusion in $A$. If we were interested in estimating the population or superpopulation mean, the Horvitz-Thompson weighted average is a standard choice, and is given by $n^{-1} \sum_{i=1}^{n} w_i^A X_i$, where $w_i^A = \alpha_i d_i$.

Given that the Horvitz-Thompson weighted average is an effective estimator of the population or superpopulation mean, it is natural to entertain the possibility that a Horvitz-Thompson weighted median is an effective estimator of the population or superpopulation median. For a given (possibly random) weight vector $w = (w_1, \ldots, w_n)$, we define the weighted median by

$$\text{med}(w) = \frac{X_{(l(w))} + X_{(u(w))}}{2},$$

where $l(w)$ and $u(w)$ are equal to the lowest index $j$ for which the cumulative weight $\sum_{i=1}^{j} w_{(i)} / \sum_{i=1}^{n} w_i$ is at least 0.5 and strictly greater than 0.5, respectively, where $w_{(i)}$ is the weight assigned to the $i^{\text{th}}$ smallest observation, $X_{(i)}$. Then the standard population median in (1) can also be expressed as a weighted median where all weights are equal to one. The Horvitz-Thompson-based survey median of Kuk [5] is given by

$$\hat{\theta}_n^A = \text{med}(w^A),$$

recalling that $w_i^A = \alpha_i d_i$.

Now suppose that we also have at our disposal observations from a big data stratum $B$, with an unknown sampling mechanism. One way to construct an integrated median $\hat{\theta}_n^{DI}$ is to adopt the weights used by Kim and Tam [6] to define

$$w_i^{DI} = \delta_i + (1 - \delta_i) w_i^A,$$

$$\hat{\theta}_n^{DI} = \text{med}(w^{DI}),$$

where $\delta_i$ equals 1 or 0 according to whether or not $i$ lies in $B$. Like Kim and Tam [6], we assume that it is possible to identify elements in $A$ that are also in $B$, so that we know $\delta_i$. We will see later that this assumption can be easily satisfied by restricting $B$ to the subset of big-data observations for which (non)membership in $A$ is known.

The remainder of the paper is dedicated to three results, which hold under mild regularity conditions. Firstly, the population, survey and integrated medians are consistent, in that they all converge to the superpopulation median $\theta_0$. This also implies that the survey and integrated medians are good approximations to the population median, in the sense that their respective distances to the population median are small for large populations. If the observations are continuous around the superpopulation median, our latter two results show that: 1) all medians considered are asymptotically unbiased, and 2) the asymptotic variance of the integrated median is less than that of the survey median. These three results reflect the ability of the integrated median to take advantage of the information in the big data sample while retaining the representativeness of the survey, and thereby produce a statistic that is a more accurate representation than its survey-only counterpart.

## 3 Consistency

The following theorem provides simple sufficient conditions ensuring that the population, survey and integrated medians are close to the superpopulation median when sampling from large populations. We will defer discussion of the impact of the survey and big-data sample sizes until Section 4.

**Theorem 3.1**
*Assume that:*

  a) *The quantile function of $X_i$ is continuous at 0.5.*
  b) *The sequence $(X_1, \pi_1, \delta_1), \ldots, (X_n, \pi_n, \delta_n)$ is i.i.d.*
  c) *The first-order inclusion probabilities $\pi_i$ are almost surely positive.*
  d) *The sequence $\alpha_1, \ldots, \alpha_n$ are i.i.d. such that $\alpha_i | X_i, \pi_i, \delta_i \sim Bernoulli(\pi_i)$.*

*Then*

$$\hat{\theta}_n \to \theta_0,$$

$$\hat{\theta}_n^A \to \theta_0,$$

$$\hat{\theta}_n^{DI} \to \theta_0,$$

*almost surely as* $n \to \infty$.

Because all medians converge to the same value, it follows immediately that

$$\left|\hat{\theta}_n^A - \hat{\theta}_n\right| \to 0,$$

$$\left|\hat{\theta}_n^{DI} - \hat{\theta}_n\right| \to 0,$$

so that the survey and integrated medians are both close to the population median for large population sizes.

Assumption a) excludes cases where there are many possible superpopulation medians $\theta_0$. For an example, consider a superpopulation given by the Bernoulli(0.5) distribution, where $\Pr(X_i \leq \theta_0) \geq 0.5$ and $\Pr(X_i \geq \theta_0) \geq 0.5$ for all $0 \leq \theta_0 \leq 1$. Violating this assumption requires a zero-density gap in the superpopulation *immediately* after its cumulative distribution function reaches *exactly* 0.5, and such specific behaviour can only realistically come about in artificially constructed scenarios.

The assumption that the data (and in this case, selection into the two strata) are i.i.d. is standard in much of the literature on summary statistics like the median and is perhaps the most questionable assumption we make. The i.i.d. assumption is more appropriate in cross-sectional contexts (e.g. Chapter 1 of Wooldridge [8]) than it is in time-series contexts (e.g.

Chapter 3.1 of Hamilton [9]) or spatial contexts (e.g. Anselin [10]), where dependence between observations can be substantial. We can relax this assumption at the cost of increased technicality if observations are not too dependent on too many of their peers (see Chapters 20, 21 and 25 of Davidson [11]).

Note that Assumption b) places no restriction on the dependence between random variables that relate to the same observation. In particular, we are permitted to exclude observations from the big-data stratum $B$ on the basis that we are uncertain about their membership in $A$, even if this would create a dependence between the big-data and survey indicators $\delta_i$ and $\alpha_i$. This provides a simple solution to the problem of uncertain linkage if we remain sure about the survey (non)membership of a substantial subset of big-data observations.

Assumption c) is a basic requirement of probability sampling and is crucial for both the survey and integrated medians. Note that we can always redefine the population to include only those units for which sampling is possible. In this case, we should be careful to interpret the two medians as being representative of the restricted population.

Finally, in d) we assume that the survey is produced via Poisson sampling. This assumption is easy to relax for sampling schemes where the size of the sample is predetermined, because this only introduces a small amount of dependence that vanishes rapidly with increasing population size.

## 4   Asymptotic Unbiasedness and Efficiency of the Integrated Median

The following theorem provides asymptotic distributions for the population, survey, and integrated medians. Here, '$\Rightarrow$' denotes convergence in distribution; see Section 25 of Billingsley [12].

**Theorem 4.1**
*Take as given the assumptions of Theorem 3.1, and assume that:*

a) $\mathbb{E}[d_i] < \infty$.

b) *The conditional density of $X_i$ given $\alpha_i = 1$ exists and is bounded away from zero in a neighbourhood of the median.*

*Then if $f$ is the (unconditional) density of $X_i$, $f(\theta_0) > 0$ and*

$$\sqrt{n}(\hat{\theta}_n - \theta_0) \Rightarrow N(0, V), \quad V = \frac{1}{4f(\theta_0)^2},$$

$$\sqrt{n}(\hat{\theta}_n^A - \theta_0) \Rightarrow N(0, V^A), \quad V^A = V + \frac{\mathbb{E}[d_i - 1]}{4f(\theta_0)^2},$$

$$\sqrt{n}(\hat{\theta}_n^{DI} - \theta_0) \Rightarrow N(0, V^{DI}), \quad V^{DI} = V^A - \frac{\mathbb{E}[\delta_i(d_i - 1)]}{4f(\theta_0)^2}.$$

There are three primary implications of this result for the relative quality of the median estimators. First, all medians are asymptotically unbiased estimators of the superpopulation median. By linearity of the expectation operator, it follows immediately that both the survey and integrated medians are asymptotically unbiased estimators of the population median, too.

Second, the asymptotic variances satisfy $V \leq V^{DI} \leq V^A$, with $V = V^{DI} = V^A$ only if the population and survey strata coincide (so that $d_i = \pi_i = 1$ for all $i$), otherwise $V = V^{DI}$ only if the population and big-data strata coincide (so that $\delta_i = 1$ for all $i$), and $V^{DI} = V^A$ only if the big-data stratum is empty (so that $\delta_i = 0$ for all $i$). With the integrated median retaining the asymptotic unbiasedness of the survey median while achieving a smaller asymptotic variance, the aim of integrating big- and survey-data to produce a more accurate but still representative estimator of the median has been realised. We see no reason to prefer the survey median over its integrated counterpart when sampling from large populations.

Finally, the asymptotic variance of the survey and integrated medians are smaller for larger selection probabilities $\pi_i$ and $\mathbb{E}[\delta_i]$, which lead to larger sample sizes for the survey and big-data strata, respectively. For the big-data median, asymptotic variance is further reduced if the survey has lower selection probabilities for units that are well-covered by the big-data stratum (and higher selection probabilities elsewhere), so that $\mathbb{E}[\delta_i d_i] = \mathbb{E}\left[\frac{\delta_i}{\pi_i}\right]$ is high (and $\mathbb{E}[d_i]$ is unchanged).

It is important to note that Assumption b) requires that the superpopulation possess a density around the median, and so this theorem does not apply to discrete data. Other restrictions imposed by the two additional assumptions are technical in nature and unlikely to be violated in real-world contexts. Assumption a) is satisfied if it is impossible for the survey probability of selection $\pi_i$ to fall below some positive lower bound, which is common in popular sampling schemes. Assumption b) strengthens Assumption a) of Theorem 3.1 to exclude superpopulations that have zero density at or immediately next to the median, and exclude survey strata produced using sampling schemes that are very unlikely to select observations close to the superpopulation median. For continuous data, violating this assumption requires such specific behaviour that it is not realistic except in artificially constructed scenarios.

## 5 Conclusion

In this paper we have constructed an integrated median that combines survey data and big data. The integrated median is asymptotically unbiased with an asymptotic variance that is less than the median produced with the survey data alone.

We see several avenues for further research. First, the integrated median was constructed under the assumption that there is no measurement error in the big data, but this is unrealistic in many scenarios. Kim and Tam [6] account for big-data measurement error in their integrated estimator of the mean using a regression approach. Could a similar approach be applied to the median, and what are the consequences for the asymptotic bias and variance of the resulting estimator relative to its survey-only counterpart?

Second, our method is only able to take advantage of units in the big data stratum for which we are certain about their (non)membership in the survey. Is it possible to produce an integrated median that allows for uncertain linkages (see Section 5 of Kim and Tam [6], for example) and what are the consequences?

Finally, Theorem 4.1 was produced using results on *m-estimators*, which are estimators that optimise an average objective function. This is a broad class that encompasses a range of linear and non-linear statistical models for cross-sectional, time-series and longitudinal data, and results on m-estimators provide a comprehensive framework for statistical inference. See Chapter 5 of van der Vaart [13] for a textbook introduction covering the i.i.d. case, and see Jacod and Sørensen [14] for extensions to time-series and longitudinal data. Could the results in this paper be extended to provide a framework for producing integrated m-estimators in general, beyond the median?

# 6  Acknowledgements

Thanks are owed to several of my colleagues at the ABS whose involvement has made this paper possible. I am grateful to Tim Cadogan-Cowper, Ric Clarke, Kristen Stone and Anders Holmberg for their reception and support when I proposed the project last year, and, alongside Qinghuan Luo, for their considered feedback and opinion on draft manuscripts without which the quality of the paper would not be at the level it is today. Guidance by Siu-Ming Tam in particular resulted in a substantially changed focus and rewrite of the paper, very much for the better.

*[14] Jacod J, Sørensen M. A Review of Asymptotic Theory of Estimating Functions. Statistical Inference for Stochastic Processes. 2018;21(2):415-34.*

*[15] David HA, Nagaraja HN. Order Statistics: Wiley; 2004.*

# 8  Appendix

Assume that all random variables are measurable mappings $\Omega \mapsto \mathbb{R}$ on the probability space $(\Omega, \mathcal{F}, P)$.

**Lemma 8.1**
*If $\pi_i > 0$ almost surely and $\alpha_1, \ldots, \alpha_n$ are i.i.d. such that $\alpha_i | Y, \pi_i \sim Bernoulli(\pi_i)$ for some random variable $Y$,*

$$\int \alpha_i d_i Y \, dP(\alpha_i | Y, \pi_i) = Y.$$

*Proof.*

Since $d_i$ is a function only of $\pi_i$,

$$\int \alpha_i d_i Y \, dP(\alpha_i | Y, \pi_i) = d_i Y \int \alpha_i \, dP(\alpha_i | Y, \pi_i)$$

$$= \pi_i^{-1} Y \pi_i$$

$$= Y,$$

where the second equality follows from the conditional independence assumption, and the second equality follows by construction of $\alpha_i$.

∎

**Lemma 8.2**
*If $\pi_i > 0$ almost surely and $\alpha_1, \ldots, \alpha_n$ are i.i.d. such that $\alpha_i | Y, \pi_i \sim Bernoulli(\pi_i)$ for some integrable random variable $Y$, then the following are also integrable:*

$$Y^A = \alpha_i d_i Y,$$
$$Y^{DI} = \delta_i Y + (1 - \delta_i) Y^A.$$

*Proof of integrability of $Y^A$.*

By the monotone convergence theorem (e.g. Theorem 16.2 of Billingsley [12]) and Lemma 8.1,

$$\int \min(\alpha_i d_i |Y|, n) \, dP(\alpha_i | Y, \pi_i) \uparrow \int \alpha_i d_i |Y| \, dP(\alpha_i | Y, \pi_i)$$
$$= |Y|.$$

Since $\min(\alpha_i d_i |Y|, n)$ is bounded it is integrable, and we have (Theorem 34.4, Billingsley [12])

$$\int \min(\alpha_i d_i |Y|, n) \, dP = \int \int \min(\alpha_i d_i |Y|, n) \, dP(\alpha_i | Y, \pi_i) \, dP.$$

Now apply the monotone convergence theorem to both sides of the equality:

$$\int \min(\alpha_i d_i |Y|, n) \, dP \uparrow \int \alpha_i d_i |Y| \, dP,$$
$$\int \int \min(\alpha_i d_i |Y|, n) \, dP(\alpha_i | Y, \pi_i) \, dP \uparrow \int |Y| \, dP.$$

Because the left-hand sides are equal, their right-hand sides are equal too, so that

$$\int |Y^A|\, dP = \int \alpha_i d_i |Y|\, dP$$
$$= \int |Y|\, dP$$
$$< \infty$$

where the inequality follows by assumption from the integrability of $Y$.

∎

*Proof of integrability of $Y^{DI}$.*

Since

$$|Y^{DI}| = \delta_i |Y| + (1 - \delta_i)|Y^A| \leq |Y| + |Y^A|,$$

we have integrability of $Y^{DI}$ from integrability of $Y$ and $Y^A$, and Theorem 16.1 (ii) of Billingsley [12].

∎

**Lemma 8.3**

*Given an integrable random variable Y, define*

$$Y^A = \alpha_i d_i Y,$$

$$Y^{DI} = \delta_i Y + (1 - \delta_i) Y^A.$$

*If $\pi_i > 0$ almost surely and $\alpha_1, \ldots, \alpha_n$ are i.i.d. such that $\alpha_i | Y, \pi_i \sim Bernoulli(\pi_i)$, then*

$$\mathbb{E}[Y^{DI}] = \mathbb{E}[Y^A] = \mathbb{E}[Y].$$

*If $d_i Y^2$ is also integrable, then*

$$\mathbb{E}[(Y^A)^2] = \mathbb{E}[d_i Y^2]$$

$$\mathbb{E}[(Y^{DI})^2] = \mathbb{E}[(Y^A)^2] - \mathbb{E}[\delta_i (d_i - 1) Y^2].$$

*Proof.*

We will start with the first moments. By assumption $Y$ is integrable, and $Y^A$ is integrable by Lemma 8.2, so we can apply the law of total expectation (e.g. Theorem 34.4 of Billingsley [12]) to give

$$\mathbb{E}[Y^A] = \int \alpha_i d_i Y \, dP$$
$$= \int \int \alpha_i d_i Y \, dP(\alpha_i | Y, \pi_i) \, dP$$
$$= \int Y \, dP$$
$$= \mathbb{E}[Y],$$

where the third equality follows from Lemma 8.1. Lemma 8.2 also gives integrability of $Y^{DI}$,

and another invocation of the law of total expectation gives

$$\begin{aligned}\mathbb{E}[Y^{DI}] &= \int \delta_i Y + (1-\delta_i) Y^A \, dP \\ &= \int \int \delta_i Y + (1-\delta_i) Y^A \, dP(\alpha_i | Y, \pi_i, \delta_i) \, dP \\ &= \int \delta_i Y + (1-\delta_i) \int Y^A \, dP(\alpha_i | Y, \pi_i, \delta_i) \, dP \\ &= \int \delta_i Y + (1-\delta_i) Y \, dP \\ &= \int Y \, dP \\ &= \mathbb{E}[Y],\end{aligned}$$

where the fourth equality follows from Lemma 8.1. For the second moment of $Y^A$, we have

$$\begin{aligned}\mathbb{E}[(Y^A)^2] &= \mathbb{E}[\alpha_i^2 d_i^2 Y^2] \\ &= \mathbb{E}[\alpha_i d_i d_i Y^2] \\ &= \mathbb{E}[d_i Y^2],\end{aligned}$$

where the second equality follows from $\alpha_i$ being one or zero almost surely, and the third equality is given by applying $d_i Y^2$ to the first-moments component of this theorem, proven above. Finally,

$$\mathbb{E}[(Y^{DI})^2] = \mathbb{E}[(\delta_i Y + (1-\delta_i)Y^A)^2]$$
$$= \mathbb{E}[\delta_i^2 Y^2 + 2\delta_i(1-\delta_i)YY^A + (1-\delta_i)^2(Y^A)^2]$$
$$= \mathbb{E}[\delta_i Y^2 + (1-\delta_i)\alpha_i d_i d_i Y^2]$$
$$= \mathbb{E}[\delta_i Y^2 + (1-\delta_i)\alpha_i d_i d_i Y^2 + \delta_i d_i Y^2 - \delta_i d_i Y^2]$$
$$= \mathbb{E}[\delta_i d_i Y^2 + (1-\delta_i)\alpha_i d_i d_i Y^2] - \mathbb{E}[\delta_i Y^2 d_i - \delta_i Y^2]$$
$$= \mathbb{E}[d_i Y^2] - \mathbb{E}[\delta_i Y^2(d_i - 1)]$$
$$= \mathbb{E}[(Y^A)^2] - \mathbb{E}[\delta_i(d_i - 1)Y^2],$$

where the third equality follows from $\delta_i$ being one or zero almost surely, and the second last equality is due to the first-moments component of this theorem applied to $d_i Y^2$.

∎

Given a sequence $Y_1, Y_2, \ldots, Y_n$ of observations with weights given by the vector $w = (w_1, \ldots, w_n)$, define

$$l(w; p) = \min\left\{j : \frac{\sum_{i=1}^{j} w_{(i)}}{\sum_{i=1}^{n} w_i} \geq p\right\},$$

$$u(w; p) = \min\left\{j : \frac{\sum_{i=1}^{j} w_{(i)}}{\sum_{i=1}^{n} w_i} > p\right\},$$

where $w_{(i)}$ is the weight of $Y_{(i)}$. Let

$$I(A) = \begin{cases} 1 & A \text{ is true} \\ 0 & A \text{ is false} \end{cases}.$$

**Theorem 8.1**
*Suppose that $(Y_1, w_1), (Y_2, w_2), \ldots, (Y_n, w_n)$ is an i.i.d. sequence of random vectors such that $w_i \geq 0$ and $\mathbb{E}[w_i I(Y_i \leq y)] = \mathbb{E}[I(Y_i \leq y)]$ for all $y$, and let $F^{-1}$ be the quantile function of the distribution of $Y_i$. Then if $F^{-1}$ is continuous at $p$,*

$$Y_{(l(w;p))} \to F^{-1}(p),$$
$$Y_{(u(w;p))} \to F^{-1}(p),$$
$$Y_{(L(w;p))} \to F^{-1}(p),$$

*almost surely as $n \to \infty$, where $L(w; p) = \min(l(w; p) + 1, n)$.*

*Proof.*

Let $l \equiv l(w; p), u \equiv u(w; p)$ and $L \equiv L(w; p)$. Define

$$F_n^l(y) = \frac{n^{-1} \sum_{i=1}^n w_i I(Y_i \leq y)}{n^{-1} \sum_{i=1}^n w_i},$$

$$F_n^v(y) = F_n^l(y) - n^{-1} F_n^l(y)(1 - F_n^l(y)),$$

$$Y_{(v)} = \inf\{y : F_n^v(y) \geq p\}.$$

By the strong law of large numbers and the mapping theorem (e.g. Theorem 19.8 of Davidson [11]), $F_n^l(y) \to F(y)$ and $F_n^v(y) \to F(y)$ as $n \to \infty$, almost surely. Lemma 21.2 of van der Vaart [13] then gives $Y_{(l)} \to F^{-1}(p)$ and $Y_{(v)} \to F^{-1}(p)$, almost surely, and the convergence of $Y_{(u)}$ and $Y_{(L)}$ follows from $Y_{(l)} \leq Y_{(u)} \leq Y_{(L)} \leq Y_{(v)}$.

∎

**Corollary 8.1**
*Under the assumptions of Theorem 8.1,*

$$(1 - \gamma_n)Y_{(l(w;p))} + \gamma_n Y_{(u(w;p))} \to F^{-1}(p)$$

*almost surely as $n \to \infty$ for all almost-surely bounded $\gamma_n$.*

*Proof.*

Let $l \equiv l(w, p)$ and $u \equiv u(w, p)$ for brevity. By the triangle inequality,

$$\left|(1 - \gamma_n)Y_{(l)} + \gamma_n Y_{(u)} - F^{-1}(p)\right| = \left|(1 - \gamma_n)\left(Y_{(l)} - F^{-1}(p)\right) + \gamma_n \left(Y_{(u)} - F^{-1}(p)\right)\right|$$
$$\leq |1 - \gamma_n|\left|Y_{(l)} - F^{-1}(p)\right| + |\gamma_n|\left|Y_{(u)} - F^{-1}(p)\right|$$
$$\to 0,$$

almost surely, where the convergence line follows from boundedness of $\gamma_n$ and Theorem 8.1.

∎

**Lemma 8.4**
*Under assumptions b) and d) of Theorem 3.1, the sequence*

$(\alpha_1, X_1, \pi_1, \delta_1), \ldots, (\alpha_n, X_n, \pi_n, \delta_n)$ *is i.i.d.*

*Proof.*

Let $A_1, A_2, \ldots, A_n$ be measurable sets, and define

$$f_i(w, x, y, z) = \begin{cases} 1 & (w, x, y, z) \in A_i \\ 0 & (w, x, y, z) \notin A_i \end{cases}.$$

We start by showing that the sequence is identically distributed. The law of total expectation gives

$$P((\alpha_i, X_i, \pi_i, \delta_i) \in A_1) = \int f_1(\alpha_i, X_i, \pi_i, \delta_i) \, dP$$
$$= \int \int f_1(\alpha_i, X_i, \pi_i, \delta_i) \, dP(\alpha_i | X_i, \pi_i, \delta_i) \, dP$$
$$= \int (1 - \pi_i) f_1(0, X_i, \pi_i, \delta_i) + \pi_i f_1(1, X_i, \pi_i, \delta_i) \, dP \, ,$$

where the final equality follows from Assumption d). Since $(X_i, \pi_i, \delta_i)$ are identically distributed, this probability is constant for all $i$, and $(\alpha_i, X_i, \pi_i, \delta_i)$ are identically distributed. Now show independence:

$$P\left(\bigcap_{i=1}^{n}(\alpha_i, X_i, \pi_i, \delta_i) \in A_i\right) = \int \prod_{i=1}^{n} f_i(\alpha_i, X_i, \pi_i, \delta_i) \, dP$$

$$= \int \int \prod_{i=1}^{n} f_i(\alpha_i, X_i, \pi_i, \delta_i) \, dP(\alpha | X, \pi, \delta) \, dP(X, \pi, \delta)$$

$$= \int \prod_{i=1}^{n} \int f_i(\alpha_i, X_i, \pi_i, \delta_i) \, dP(\alpha_i | \pi_i) \, dP(X, \pi, \delta)$$

$$= \prod_{i=1}^{n} \int \int f_i(\alpha_i, X_i, \pi_i, \delta_i) \, dP(\alpha_i | \pi_i) \, dP(X, \pi, \delta)$$

$$= \prod_{i=1}^{n} \int f_i(\alpha_i, X_i, \pi_i, \delta_i) \, dP$$

$$= \prod_{i=1}^{n} P\bigl((\alpha_i, X_i, \pi_i, \delta_i) \in A_i\bigr),$$

where the second and fifth equalities follow from the law of total expectation, the third equality follows from Assumption d) of Theorem 3.1, and the fourth equality follows from independence of $\pi_1, \ldots, \pi_n$.

∎

*Proof of Theorem 3.1.*

For each of these three estimators, we will prove convergence by satisfying the assumptions of Theorem 8.1 to apply Corollary 8.1. The continuity of the quantile function at $p = 0.5$ is provided by Assumption a). By construction, we also have that in all cases $\gamma_n = 0.5$, and for $\hat{\theta}_n$ there is nothing more to prove since the corresponding weights are constant. By Lemma 8.4, the sequence $(\alpha_1, X_1, \pi_1, \delta_1), \ldots, (\alpha_n, X_n, \pi_n, \delta_n)$ is i.i.d., and it follows that the sequences $(X_1, w_1^A), \ldots, (X_n, w_n^A)$ and $(X_1, w_1^{DI}), \ldots, (X_n, w_n^{DI})$ are i.i.d. as well. Assumptions c) and d) can then be applied to invoke Lemma 8.3 and obtain $\mathbb{E}[w_i^A I(X_i \leq x)] = \mathbb{E}[w_i^{DI} I(X_i \leq x)] = \mathbb{E}[I(X_i \leq x)]$ for all $x$, and all assumptions of Theorem 8.1 are now satisfied for convergence of both $\hat{\theta}_n^A$ and $\hat{\theta}_n^{DI}$.

∎

Given random variables $Y_1, \ldots, Y_n$, define

$$m_\theta(y, w; p) = w\big((1-p)I(y \leq \theta) - pI(y > \theta)\big)(y - \theta),$$

$$M_n(\theta) = \frac{1}{n} \sum_{i=1}^n m_\theta(Y_i, w_i; p).$$

**Lemma 8.5**
*If $w_i \geq 0$ for all $i$, the function $M_n(\theta)$ is maximised at $\theta = Y_{(l(w;p))}$.*

*Proof.*

If $\theta \leq \theta + \Delta \leq Y_1$, then

$$M_n(\theta + \Delta) = \frac{1}{n} \sum_{i=1}^n w_i p(\theta + \Delta - Y_i)$$
$$= \frac{1}{n} \sum_{i=1}^n w_i p(\theta - Y_i) + \frac{\Delta p}{n} \sum_{i=1}^n w_i$$
$$= M_n(\theta) + \frac{\Delta p}{n} \sum_{i=1}^n w_i,$$

so that by nonnegativity of the weights, $M_n(\theta + \Delta) - M_n(\theta) \geq 0$, and $M_n$ is nondecreasing on $(-\infty, Y_1]$. By a similar argument, $M_n$ is nonincreasing on $[Y_n, \infty)$. If $Y_j \leq \theta \leq \theta + \Delta \leq Y_{j+1}$, then

$$M_n(\theta + \Delta) = \frac{1}{n} \sum_{i=1}^n w_i \big((1-p)I(Y_i \leq \theta) - pI(Y_i > \theta)\big)(Y_i - \theta - \Delta)$$
$$= M_n(\theta) - \frac{1}{n} \sum_{i=1}^n w_i \big((1-p)I(Y_i \leq \theta) - pI(Y_i > \theta)\big)\Delta$$
$$= M_n(\theta) - \frac{1}{n} \sum_{i=1}^n w_i (I(Y_i \leq \theta) - p)\Delta$$
$$= M_n(\theta) + \Delta \left( p \frac{1}{n} \sum_{i=1}^n w_i - \frac{1}{n} \sum_{i=1}^j w_{(i)} \right),$$

so that on the interval $[Y_j, Y_{j+1}]$, $M_n$ is nondecreasing if $\sum_{i=1}^j w_{(i)} / \sum_{i=1}^n w_i < p$, $M_n$ is constant if $\sum_{i=1}^j w_{(i)} / \sum_{i=1}^n w_i = p$, and $M_n$ is nonincreasing if $\sum_{i=1}^j w_{(i)} / \sum_{i=1}^n w_i > p$. Thus $Y_{(l(w;p))}$ maximises $M_n(\theta)$.

**Lemma 8.6**

*Take as given the assumptions of Theorem 8.1, and suppose that the density of $Y_i$ conditional on $w_i > 0$ exists and is bounded away from zero in a neighbourhood of $F^{-1}(p)$. Then the (unconditional) density of $Y_i$ is bounded away from zero in the same neighbourhood, and*

$$n\left(\sup_\theta M_n(\theta) - M_n\left((1-\gamma_n)Y_{(l(w;p))} + \gamma_n Y_{(u(w;p))}\right)\right) \xrightarrow{p} 0$$

*as $n \to \infty$ for any sequence of random variables $\gamma_n \in [0,1]$.*

*Proof.*

The density $f$ of $Y_i$ is bounded away from zero because $f(y_i) = f(y_i|w_i = 0)p(w_i = 0) + p(y_i|w_i > 0)p(w_i > 0) \geq p(y_i|w_i > 0)p(w_i > 0) > 0$ by assumption. Let $l \equiv l(w;p)$, $u \equiv u(w;p)$ and $L \equiv L(w;p)$ for brevity. Then,

$$M_n\left((1-\gamma_n)Y_{(l)} + \gamma_n Y_{(u)}\right) = \frac{1}{n}\sum_{i=1}^n w_i\left(I(Y_i \leq Y_{(l)}) - p\right)\left(Y_i - (1-\gamma_n)Y_{(l)} - \gamma_n Y_{(u)}\right)$$

$$= M_n(Y_{(l)}) - \frac{1}{n}\sum_{i=1}^n w_i\left(I(Y_i \leq Y_{(l)}) - p\right)\gamma_n(Y_{(u)} - Y_{(l)})$$

$$= \sup_\theta M_n(\theta) - \gamma_n(Y_{(u)} - Y_{(l)})\left(\frac{1}{n}\sum_{i=1}^n w_i I(Y_i \leq Y_{(l)}) - p\frac{1}{n}\sum_{i=1}^n w_i\right),$$

$$n\left(\sup_\theta M_n(\theta) - M_n\left(\gamma_n Y_{(l)} + (1-\gamma_n)Y_{(u)}\right)\right) \leq n(Y_{(u)} - Y_{(l)})\left(\frac{1}{n}\sum_{i=1}^n w_i I(Y_i \leq Y_{(l)}) - p\frac{1}{n}\sum_{i=1}^n w_i\right),$$

where $M_n(Y_{(l)}) = \sup_\theta M_n(\theta)$ by Lemma 8.5. We can show that

$$\sup_t \left|\frac{1}{n}\sum_{i=1}^n w_i I(Y_i \leq t) - F(t)\right| \to 0$$

almost surely as $n \to \infty$ by replacing $\mathbb{F}_n(t)$ with $n^{-1}\sum_{i=1}^{n} w_i I(Y_i \leq t)$ in the proof of Theorem 19.1 in van der Vaart [13], and $n^{-1}\sum_{i=1}^{n} w_i I(Y_i \leq t) \to p$ follows from continuity of $F$ at $F^{-1}(p)$. By applying the monotone convergence theorem to $\mathbb{E}[w_i I(Y_i \leq k)]$ for $k \to \infty$, $\mathbb{E}[w_i] = 1$ and the strong law of large numbers gives $n^{-1}\sum_{i=1}^{n} w_i \to 1$. We will complete the proof of Lemma 8.6 by showing that $n(Y_{(u)} - Y_{(l)})$ is bounded in probability. Let $Y'_1, Y'_2, \ldots, Y'_{n'}$ be the $n'$ observations with positive weight and define $l'$ and $L'$ such that $Y'_{(l')} = Y_{(l)}$ and $Y'_{(L')} = Y_{(L)}$. Without loss of generality, suppose that $Y'_i = F^{-1}_{Y|W>0}(U_i)$ where $F^{-1}_{Y|W>0}$ is the quantile function of the positively weighted $Y_i$ and $U_1, U_2, \cdots, U_{n'}$ are i.i.d. Uniform(0,1) random variables. Then,

$$\begin{aligned}
n(Y_{(u)} - Y_{(l)}) &\leq n\left(Y'_{(L')} - Y'_{(l')}\right) \\
&= n\left(F^{-1}_{Y|W>0}\left(U_{(L')}\right) - F^{-1}_{Y|W>0}\left(U_{(l')}\right)\right) \\
&= n\, f_{Y|W>0}\left(F^{-1}_{Y|W>0}(\tilde{U})\right)^{-1}\left(U_{(L')} - U_{(l')}\right) \\
&= n f_{Y|W>0}(\tilde{Y})^{-1}\left(U_{(L')} - U_{(l')}\right),
\end{aligned}$$

where the second equality follows from the mean value theorem: $U_{(l')} \leq \tilde{U} \leq U_{(L')}$, $\tilde{Y} = F^{-1}_{Y|W>0}(\tilde{U})$, and $f_{Y|W>0}(\tilde{Y})^{-1}$ is the derivative of $F^{-1}_{Y|W>0}$ evaluated at $\tilde{U}$ by Lemma 21.1 (ii) of van der Vaart [13]. This derivative is bounded for large $n$ by virtue of $f_{Y|W>0}$ being bounded away from zero in a neighbourhood of $F^{-1}(p)$, and convergence of $\tilde{Y}$ to $F^{-1}(p)$ is implied by $Y_{(l)} \leq \tilde{Y} \leq Y_{(L')}$ and Theorem 8.1. Finally, if $l' < n'$, $U_{(l'+1)} - U_{(l')}$ has a Beta$(1, n+1)$ distribution (e.g., Example 2.3 of David and Nagaraja [15]), and we have

$$\mathbb{E}\left[\left|n\left(U_{(L')} - U_{(l')}\right)\right|\right] \leq \frac{n'}{n'+1} < 1.$$

Therefore $n\left(U_{(L')} - U_{(l')}\right)$ is bounded in probability by Markov's inequality, and by the mapping theorem $n\left(\sup_\theta M_n(\theta) - M_n\left((1-\gamma_n)Y_{(l)} + \gamma_n Y_{(u)}\right)\right)$ converges to zero in probability.

∎

**Theorem 8.2**
*Take as given the assumptions of Theorem 3.1. Let $\vartheta_0 = F^{-1}(p)$ where $F^{-1}$ is the quantile function of $Y_i$ and assume that:*

a) $\mathbb{E}[d_i] < \infty$.

b) *The conditional density of $Y_i$ given $\alpha_i = 1$ exist and is bounded away from zero in a neighbourhood of $\vartheta_0$.*

*Take any sequence of random variables $\gamma_n \in [0,1]$, and define*

$$\hat{\vartheta}_n = (1-\gamma_n)Y_{(l(1;p))} + \gamma_n Y_{(u(1;p))},$$
$$\hat{\vartheta}_n^A = (1-\gamma_n)Y_{(l(w^A;p))} + \gamma_n Y_{(u(w^A;p))},$$
$$\hat{\vartheta}_n^{DI} = (1-\gamma_n)Y_{(l(w^{DI};p))} + \gamma_n Y_{(u(w^{DI};p))}.$$

*Then if $f$ is the (unconditional) density of $Y_i$, $f(\vartheta_0) > 0$, and we have*

$$\sqrt{n}(\hat{\vartheta}_n - \vartheta_0) \Rightarrow N(0,V), \quad V = \frac{p(1-p)}{4f(\vartheta_0)^2},$$
$$\sqrt{n}(\hat{\vartheta}_n^A - \vartheta_0) \Rightarrow N(0,V^A), \quad V^A = V + \Delta^A,$$
$$\sqrt{n}(\hat{\vartheta}_n^{DI} - \vartheta_0) \Rightarrow N(0,V^{DI}), \quad V^{DI} = V^A - \Delta^{DI},$$

$$\Delta^A = \frac{(1-p)^2 \mathbb{E}[(d_i-1)I(Y_i \leq \vartheta_0)] + p^2 \mathbb{E}[(d_i-1)I(Y_i > \vartheta_0)]}{f(\vartheta_0)^2},$$

$$\Delta^{DI} = \frac{(1-p)^2 \mathbb{E}[\delta_i(d_i-1)I(Y_i \leq \vartheta_0)] + p^2 \mathbb{E}[\delta_i(d_i-1)I(Y_i > \vartheta_0)]}{f(\vartheta_0)^2}.$$

*Proof.*

All convergence results will be obtained by satisfying the assumptions of Theorem 5.23 of van der Vaart [13]. Because $f$ is continuous at $\vartheta_0$, $\vartheta \mapsto m_\vartheta(y, w; p)$ is almost-surely differentiable, despite no derivative existing when $y = \vartheta_0$. Where it exists, the derivative is given by

$$\dot{m}_\vartheta(y, w; p) = -w\big((1-p)I(y \leq \vartheta) - pI(y > \vartheta)\big).$$

Take any $\vartheta_1$ and $\vartheta_2$ in a neighbourhood of $\vartheta_0$. It is easy to show that

$$m_{\vartheta_2}(y, w; p) = m_{\vartheta_1}(y, w; p) + \int_{\vartheta_1}^{\vartheta_2} \dot{m}_\vartheta(y, w; p)\, d\vartheta,$$

and therefore $\vartheta \mapsto m_\vartheta(y, w; p)$ is absolutely continuous, and we have

$$\big|m_{\vartheta_1}(y, w; p) - m_{\vartheta_2}(y, w; p)\big| \leq \sup_\vartheta |\dot{m}_\vartheta(y, w; p)|\, |\vartheta_1 - \vartheta_2|$$

$$= w \max(p, 1-p)\, |\vartheta_1 - \vartheta_2|$$

$$\leq w|\vartheta_1 - \vartheta_2|.$$

For $\hat{\vartheta}_n$, we have $w = 1$, so that $\dot{m}(y, w; p) = 1$ suffices (for the role of $\dot{m}(y, w; p)$ see

Theorem 5.23 of van der Vaart [13]). By Lemma 8.3, $\mathbb{E}[w_i^{DI}] = \mathbb{E}[w_i^A] = 1$, so $\dot{m}(y, w; p) = w_i^A$ and $\dot{m}(y, w; p) = w_i^{DI}$ suffices for $\hat{\vartheta}_n^A$ and $\hat{\vartheta}_n^{DI}$, respectively. Now take the expectation of $m_\vartheta$:

$$\mathbb{E}[m_\vartheta(Y_i, w; p)] = \mathbb{E}[w(I(Y_i \leq \vartheta) - p)(Y_i - \vartheta)]$$
$$= \mathbb{E}[(I(Y_i \leq \vartheta) - p)(Y_i - \vartheta)]$$
$$= \mathbb{E}[I(Y_i \leq \vartheta) Y_i] - \vartheta F(\vartheta) + p(\vartheta - \mathbb{E}[Y_i]),$$

where any of $w = 1$, $w = w^A$, or $w = w^{DI}$ works by Lemma 8.3. The expectation $\mathbb{E}[m_\vartheta(Y_i, w; p)]$ is twice differentiable with derivatives

$$\left.\frac{\partial \mathbb{E}[m_\vartheta(Y_i, w; p)]}{\partial \vartheta}\right|_{\vartheta = \vartheta_0} = \vartheta_0 f(\vartheta_0) - F(\vartheta_0) - \vartheta_0 f(\vartheta_0) + p$$
$$= -F(\vartheta_0) + p,$$
$$\left.\frac{\partial^2 \mathbb{E}[m_{\vartheta_0}(Y_i, w; p)]}{\partial \vartheta_0^2}\right|_{\vartheta = \vartheta_0} = -f(\vartheta_0),$$

so $\vartheta \mapsto \mathbb{E}[m_\vartheta(Y_i, w; p)]$ admits a second-order Taylor expansion at a point of maximum $\vartheta_0$, where the second derivative is non-zero by Lemma 8.6. The result of evaluating $M_n$ at any of the three estimators lies within $o_p(n^{-1})$ of $\sup_\vartheta M_n(\vartheta)$ by Lemma 8.6, and Corollary 8.1 provides for convergence to $\vartheta_0$ since continuity of $F^{-1}$ at $p$ follows from $f$ being bounded away from zero at $\vartheta_0$. Having satisfied all its assumptions, all that remains is to simplify the expressions for $V$, $V^A$ and $V^{DI}$ given by Theorem 5.23 of van der Vaart [13], applying Lemma 8.3 as required:

$$\dot{m}_\vartheta^2(y,w;p) = w^2\big((1-p)^2 I(y \leq \vartheta) + p^2 I(y > \vartheta)\big),$$

$$\mathbb{E}[\dot{m}_{\vartheta_0}^2(Y_i,1;p)] = (1-p)^2 \mathbb{E}[I(Y_i \leq \vartheta_0)] + p^2 \mathbb{E}[I(Y_i > \vartheta_0)]$$

$$= (1-p)^2 p + p^2(1-p)$$

$$= p(1-p),$$

$$\mathbb{E}[\dot{m}_{\vartheta_0}^2(Y_i,w^A;p)] = \mathbb{E}[(w^A \dot{m}_{\vartheta_0}(Y_i,1;p))^2]$$

$$= \mathbb{E}[d_i \dot{m}_{\vartheta_0}^2(Y_i,1;p)]$$

$$= (1-p)^2 \mathbb{E}[d_i I(Y_i \leq \vartheta_0)] + p^2 \mathbb{E}[d_i I(Y_i > \vartheta_0)]$$

$$= \mathbb{E}[\dot{m}_{\vartheta_0}^2(Y_i,1;p)] + (1-p)^2 \mathbb{E}[(d_i - 1)I(Y_i \leq \vartheta_0)] + p^2 \mathbb{E}[(d_i - 1)I(Y_i > \vartheta_0)],$$

$$\mathbb{E}[\dot{m}_{\vartheta_0}^2(Y_i,w^{DI};p)] = \mathbb{E}[(w^{DI} \dot{m}_{\vartheta_0}(Y_i,1;p))^2]$$

$$= \mathbb{E}[(w^A \dot{m}_{\vartheta_0}(Y_i,1;p))^2] - \mathbb{E}[\delta_i(d_i - 1)\dot{m}_{\vartheta_0}^2(Y_i,1;p)],$$

$$= \mathbb{E}[(w^A \dot{m}_{\vartheta_0}(Y_i,1;p))^2]$$

$$- (1-p)^2 \mathbb{E}[\delta_i(d_i - 1)I(Y_i \leq \vartheta_0)] - p^2 \mathbb{E}[\delta_i(d_i - 1)I(Y_i > \vartheta_0)],$$

$$V = \frac{\mathbb{E}[\dot{m}_{\vartheta_0}^2(Y_i,1;p)]}{\left(\frac{\partial^2 \mathbb{E}[m_\vartheta(Y_i,1;p)]}{\partial \vartheta^2}\Big|_{\vartheta=\vartheta_0}\right)^2}$$

$$= \frac{p(1-p)}{f(\vartheta_0)^2},$$

$$V^A = \frac{\mathbb{E}[\dot{m}_{\vartheta_0}^2(Y_i,w^A;p)]}{\left(\frac{\partial^2 \mathbb{E}[m_\vartheta(Y_i,w^A;p)]}{\partial \vartheta^2}\Big|_{\vartheta=\vartheta_0}\right)^2}$$

$$= \frac{\mathbb{E}[\dot{m}_{\vartheta_0}^2(Y_i,1;p)] + (1-p)^2 \mathbb{E}[(d_i - 1)I(Y_i \leq \vartheta_0)] + p^2 \mathbb{E}[(d_i - 1)I(Y_i > \vartheta_0)]}{f(\vartheta_0)^2}$$

$$= V + \frac{(1-p)^2 \mathbb{E}[(d_i - 1)I(Y_i \leq \vartheta_0)] + p^2 \mathbb{E}[(d_i - 1)I(Y_i > \vartheta_0)]}{f(\vartheta_0)^2},$$

$$V^{DI} = \frac{\mathbb{E}[\dot{m}_{\vartheta_0}^2(Y_i,w^{DI};p)]}{\left(\frac{\partial^2 \mathbb{E}[m_\vartheta(Y_i,w^{DI};p)]}{\partial \vartheta^2}\Big|_{\vartheta=\vartheta_0}\right)^2}$$

$$= \frac{\mathbb{E}[(w^A \dot{m}_{\vartheta_0}(Y_i,1;p))^2] - (1-p)^2 \mathbb{E}[\delta_i(d_i-1)I(Y_i \leq \vartheta_0)] - p^2 \mathbb{E}[\delta_i(d_i-1)I(Y_i > \vartheta_0)]}{f(\vartheta_0)^2}$$

$$= V^A - \frac{(1-p)^2 \mathbb{E}[\delta_i(d_i - 1)I(Y_i \leq \vartheta_0)] + p^2 \mathbb{E}[\delta_i(d_i - 1)I(Y_i > \vartheta_0)]}{f(\vartheta_0)^2}.$$

∎

*Proof of Theorem 4.1.*

Theorem 4.1 is a restatement of Theorem 8.2 restricted to the case $\gamma_n = p = 0.5$.

∎